\documentstyle[preprint,aps,epsf]{revtex}
\begin{document}
\newcommand{\be}{\begin{equation}}
\newcommand{\ee}{\end{equation}}
\newcommand{\bea}{\begin{eqnarray}}
\newcommand{\eea}{\end{eqnarray}}
\title{Fisher zeros of the $Q$-state Potts model 
in the complex temperature plane for nonzero external magnetic field}
\author{Seung-Yeon Kim\footnote{Electronic address: kim@cosmos.psc.sc.edu}
and Richard J. Creswick\footnote{Electronic address: creswick.rj@sc.edu}}
\address{Department of Physics and Astronomy,
University of South Carolina,\\ Columbia, South Carolina 29208}
\maketitle

\begin{abstract}
The microcanonical transfer matrix is used to study the distribution
of the Fisher zeros of the $Q>2$ Potts models in the complex 
temperature plane with nonzero external magnetic field $H_q$. 
Unlike the Ising model for $H_q\ne0$
which has only a non-physical critical point (the Fisher edge singularity),
the $Q>2$ Potts models have physical critical points for $H_q<0$
as well as the Fisher edge singularities for $H_q>0$.
For $H_q<0$ the cross-over of the Fisher zeros of 
the $Q$-state Potts model into those of the ($Q-1$)-state Potts model 
is discussed, and the critical line of the three-state Potts ferromagnet
is determined.
For $H_q>0$ we investigate the edge singularity for finite lattices
and compare our results with
high-field, low-temperature series expansion of Enting.
For $3\le Q\le6$ we find that the specific heat, magnetization, 
susceptibility, and the density of zeros diverge at the Fisher
edge singularity with exponents $\alpha_e$, $\beta_e$, and $\gamma_e$
which satisfy the scaling law $\alpha_e+2\beta_e+\gamma_e=2$. 
\end{abstract}
\pacs{PACS number(s): 05.50.+q, 05.70.$-$a, 64.60.Cn, 75.10.Hk}


\section{introduction}

The $Q$-state Potts model\cite{potts,wu} in two dimensions
exhibits a rich variety of critical behavior
and is very fertile ground for the analytical and numerical
investigation of first- and second-order phase transitions.
With the exception of the $Q=2$ Potts (Ising) model
in the absence of an external magnetic field\cite{onsager}, 
exact solutions for arbitrary $Q$ are not known.
However, some exact results have been established for the
$Q$-state Potts model.
For $Q=2$, 3 and 4 there is
a second-order phase transition, while for $Q>4$
the transition is first order\cite{baxter1}.
From the duality relation the critical temperature
is known to be $T_c=J/k_B{\rm ln}(1+\sqrt{Q})$\cite{potts}.
For $Q=3$ and 4 the critical exponents\cite{dennijs}
are known, while for $Q > 4$ the latent heat\cite{baxter1}, 
spontaneous magnetization\cite{baxter2},
and correlation length\cite{buffernoir} at $T_c$ are also known.                                                                        

By introducing the concept of the zeros of the partition function
in the {\it complex} magnetic-field plane (Yang-Lee zeros),
Yang and Lee\cite{yang} proposed a mechanism
for the occurrence of phase transitions in the thermodynamic limit
and yielded a new insight into the unsolved problem of the Ising model
in an arbitrary nonzero external magnetic field. Lee and Yang\cite{yang} also 
formulated the celebrated circle theorem which states that
the Yang-Lee zeros of the Ising ferromagnet lie on the unit circle
$x_0=e^{i\theta}$ in the complex $x={\rm exp}(\beta H)$ plane
for any size lattice and any type of boundary conditions.
The density of zeros contains all the information
about a system and in particular in the thermodynamic limit the 
density of zeros completely determine the critical behavior 
of the system\cite{yang,abe,creswick2}.  
For example, the spontaneous magnetization of the Ising model
is determined by the density of zeros on the positive real axis,
i.e., $m_0(T)=2\pi g(\theta=0,T)$.
Above the critical temperature $T_c$, there is a gap in the distribution
of zeros, centered at $\theta=0$, that is, $g(\theta,T)=0$ for
$|\theta| < \theta_0(T)$. Within this gap the free energy is analytic
and there is no phase transition. The Yang-Lee zeros at 
$\theta=\pm\theta_0$ are called the Yang-Lee edge zeros.
As $T-T_c\to0^+$, $\theta_0(T)\to0$.
At $T_c$ the gap disappears, i.e., $\theta_0(T_c)=0$, and $g(0,T_c)=0$,
which is the characteristic of a second-order phase transition.
Below $T_c$, $g(0,T) > 0$ and we have a finite spontaneous
magnetization. Kortman and Griffiths\cite{kortman} carried out 
the first systematic investigation
of $g(\theta,T)$, based on the high-field, high-temperature
series expansion for the Ising model on the square lattice and
the diamond lattice. They found that above $T_c$, $g(\theta,T)$
diverges at $\theta=\pm\theta_0$, i.e., at the Yang-Lee edge singularity
for the square lattice. The divergence of the density of the Yang-Lee
zeros means the magentization diverges, which does not occur
at a physical critical point. Fisher\cite{fisher1} proposed the idea that the 
singularity at the Yang-Lee edge can be thought of as 
a new second-order phase transition  with associated critical exponents 
and the Yang-Lee edge zero can be considered
as a conventional critical point. 
Fisher also renamed the Yang-Lee edge zero as the 
Yang-Lee edge {\it singularity}. 
The critical point of the Yang-Lee edge singularity
is associated with a $\phi^3$ theory, different from the usual
critical point associated with the $\phi^4$ theory.
The crossover dimension of the Yang-Lee edge singularity is $d_c=6$.   
The study of the Yang-Lee edge singularity has been extended to 
the classical $n$-vector model\cite{kurtze1}, 
the quantum Heisenberg model\cite{kurtze1},
the spherical model\cite{kurtze2}, 
the quantum one-dimensional transverse Ising model\cite{uzelac}, 
the hierarchical model\cite{baker}, 
and the one-dimensional Potts model\cite{mittag}. 
Using Fisher's idea and conformal field theory,
Cardy \cite{cardy} studied the Yang-Lee edge singularity 
for a two-dimensional $\phi^3$ theory.
Recently the Yang-Lee zeros of the two-dimensional $Q$-state Potts
model have been studied\cite{kim}.

In 1964 Fisher\cite{fisher2} emphasized that the partition
function zeros in the complex temperature plane (Fisher zeros)
is also very useful in understanding phase transitions.
In particular, in the complex temperature plane both the ferromagnetic
phase and the antiferromagnetic phase can be considered
at the same time.
From the exact solutions\cite{onsager} of the square lattice Ising model
Fisher conjectured that in the absence of an external magnetic field 
the zeros of the partition function 
lie on two circles in the complex $y={\rm exp}(-\beta J)$ plane
given by $y_{FM}=-1+\sqrt{2}e^{i\theta}$ (ferromagnetic circle)
and $y_{AFM}=1+\sqrt{2}e^{i\theta}$ (antiferromagnetic circle).
Fisher also showed that the logarithmically infinite specific heat
singularity of the Ising model results from the properties of the 
density of zeros.
By numerical investigations\cite{katsura} and 
analytical methods\cite{brascamp} it has been concluded that
for very special boundary conditions the Fisher zeros of the Ising model
do indeed lie on two circles, while for more general boundary conditions
the zeros approach two circles as the size of lattices increases.
Recently the locus of the Fisher zeros of the $Q$-state Potts model
in the absence of an external magnetic field
has been studied extensively\cite{martin1,bhanot,martin2,shrock2,creswick3}.
It has been shown\cite{martin2}
that for self-dual boundary conditions 
near the ferromagnetic critical point $y_c=1/(1+\sqrt{Q})$
the Fisher zeros of the Potts model on a finite square lattice
lie on the circle with center $-1/(Q-1)$ and radius $\sqrt{Q}/(Q-1)$
in the complex $y$-plane, while the antiferromagnetic circle 
of the Ising model completely disappears for $Q>2$.
It is also known\cite{martin2} that all the Fisher zeros of the one-state 
Potts model lie at $y^{-1}=0$.
Shrock {\it et al.} showed that for the two-dimensional 
Ising\cite{shrock1} and Potts\cite{shrock2} models
in the absence of an external magnetic field there exist non-physical 
critical points in the complex temperature plane,
at which thermodynamic functions including the magnetization diverge. 
Itzykson {\it et al.}\cite{itzykson} considered the Fisher zeros
in an external magnetic field for the first time.
They studied the movement of the Fisher zero closest to the positive
real axis for the Ising model as the strength of a magnetic field
changes. For nonzero magnetic field there is a gap in the distribution
of the Fisher zeros of the Ising model around the positive real axis 
even in the thermodynamic limit, which means that there is no
phase transition. 
Matveev and Shrock\cite{matveev} studied the Fisher zeros of the 
two-dimensional Ising model in an external magnetic field 
using the high-field, low-temperature series expansion 
and the partition functions of finite-size systems.
They found that for nonzero magnetic field the magnetization, 
susceptibility, specific heat, and the density of zeros 
diverge at the Fisher zero closest to the 
positive real axis, which we call the Fisher edge singularity. 
In this paper we discuss the Fisher zeros of the $Q$-state Potts model
for nonzero magnetic field using the microcanonical transfer matrix
and the high-field, low-temperature series expansion.


\section{microcanonical transfer matrix and symmetries}

The $Q$-state Potts model on a lattice $G$ 
in an external magnetic field $H_q$
is defined by the Hamiltonian
\be
{\cal H}_Q=-J\sum_{<i,j>} \delta(\sigma_i,\sigma_j)-H_q\sum_k\delta(\sigma_k,q),
\ee
where $J$ is the coupling constant, 
$<i,j>$ indicates a sum over nearest-neighbor pairs,
$\sigma_i=0,...,Q-1$,
and $q$ is a fixed integer between 0 and $Q-1$.
The partition function of the model is
\be
Z_Q=\sum_{\{ \sigma_n \}} e^{-\beta{\cal H}_Q},
\ee
where $\{ \sigma_n\}$ denotes a sum over all possible configurations
and $\beta=(k_B T)^{-1}$.
The partition function can be written as 
\be 
Z_Q=y^{-N_b}\sum_{E=0}^{N_b}\sum_{M=0}^{N_s}\Omega_Q(E,M) x^M y^E,
\ee
where $x=e^{\beta H_q}$, $y=e^{-\beta J}$,
$E$ and $M$ are positive integers $0\le E\le N_b$ and $0\le M\le N_s$,
respectively, $N_b$ and $N_s$ are the number of bonds
and the number of sites on the lattice $G$,
and $\Omega_Q(E,M)$ is the number of states
with fixed $E$ and fixed $M$.
Using the microcanonical transfer matrix ($\mu$TM)
\cite{creswick1,creswick2,creswick3,kim}
we have calculated the number of states $\Omega_Q(E,M)$
of the $Q$-state Potts model on finite square lattices with
self-dual boundary conditions\cite{martin2} and 
cylindrical boundary conditions for $3\le Q\le8$.

In the absence of an external magnetic field
the partition function of the $Q$-state Potts model 
is symmetric under the dual transformation
\be
y\to{1-y\over1+(Q-1)y},
\ee
which gives the critical point
\be
y_c={1\over1+\sqrt{Q}}
\ee
and the invariant ferromagnetic circle of the Fisher zeros
\be
y_0(\theta)={-1+\sqrt{Q}e^{i\theta}\over Q-1}.
\ee
The partition function of the Ising model has the additional symmetry
\be
y\to{1\over y},
\ee
which maps the ferromagnetic Ising model to the antiferromagnetic model.
This, together with the dual transformation, implies the invariance of 
the antiferromagnetic circle 
\be
y_0=1+\sqrt{2}e^{i\theta}.
\ee
However, the $Q > 2$ Potts models do not possess this second
symmetry and the associated Fisher zeros 
are scattered in the non-critical region.
For nonzero magnetic field the Ising model also has the symmetry 
\be
x\to{1\over x},
\ee
and the Fisher zeros for $0 < x < 1$ have the same properties
as those for $1 < x < \infty$.
The $Q > 2$ Potts models do not have this symmetry,
and distribution of zeros for $0 < x < 1$ is different from
the distribution for $1 < x < \infty$.
Because the $Q > 2$ Potts models are less symmetric than 
the Ising model, the zeros of the partition function have 
a much richer structure.
For example, the Ising model has only non-physical critical points
in the complex $y$-plane for $x\ne1$, while the $Q > 2$ Potts models
have both non-physical and physical critical points 
in the same plane for $x\ne1$.
In this paper we study the Fisher zeros of the $Q$-state Potts model
for nonzero magnetic field to unveil some of the rich structures of the model.


\section{Fisher zeros of the three-state Potts model for $\lowercase{x}<1$}

In the limit $H_q\to-\infty$ ($x\to0$) the partition function 
of the $Q$-state Potts model becomes
\be 
Z_Q=y^{-N_b}\sum_{E=0}^{N_b}\Omega_Q(E,M=0) y^E,
\ee
where $\Omega_Q(E,M=0)$ is the same as the number of states 
$\Omega_{Q-1}(E)$ of the ($Q-1$)-state Potts model 
in the absence of an external magnetic field.
As $x$ decreases from 1 to 0, the $Q$-state Potts model is transformed
into the ($Q-1$)-state Potts model in zero external field. 
Figure 1 shows the Fisher zeros in the complex $y$-plane 
of the three-state Potts model for $x\le1$ with self-dual
boundary conditions. Note that in the absence of an external 
magnetic field for self-dual boundary conditions
the Fisher zeros in the critical region of the Potts model 
lie on the circle given by Eq. (6)\cite{martin2}.
In Figure 1 (a) the circle is that of Eq. (6) with $Q=3$
(the three-state Potts circle), while in Figures 1 (c) and 1 (d)
the circle is for $Q=2$ (the Ising circle).
In Figure 1 (b) we show both the the three-state Potts circle
(smaller one) and the Ising circle (larger one),
and the Fisher zeros lie on neither the three-state Potts circle
nor the Ising circle.
In Figure 1 (c) the Fisher zeros near the ferromagnetic critical
point begin to approach the Ising circle, 
and the antiferromagnetic circle of the Ising model begins to appear.
In Figure 1 (d) almost all of the Fisher zeros, which will ultimately 
lie on the ferromagnetic circle of the Ising model at $x=0$, 
are very close to this locus,
and the antiferromagnetic circle becomes clearer. 


\section{critical point of the three-state Potts model in 
a field $H_{\lowercase{q}}<0$}

For an external field $H_q<0$, one of the Potts states is supressed
relative to the others. The symmetry of the Hamiltonian is that of the
$(Q-1)$-state Potts model in zero external field, so that we expect
to see cross-over from the $Q$-state critical point to the $(Q-1)$-state
critical point as $-H_q$ is increased.

We have studied the field dependence of the critical point for 
$0\le x\le1$ through the Fisher zero closest to the real axis, $y_1(x,L)$.
For a given applied field $y_1$ approaches the critical point
$y_1(L)\to y_c(x)$ in the limit $L\to\infty$, and the thermal 
exponent $y_t(L)$ defined as\cite{bhanot,itzykson}
\be
y_t(L)=-{{\rm ln \{Im}[y_1(L+1)]/{\rm Im}[y_1(L)]\}\over{\rm ln}[(L+1)/L]}
\ee
will approach the critical exponent $y_t(x)$.
Table I shows values for $y_c(x)$ extrapolated from calculations of 
$y_1(x,L)$ on $L\times L$ lattices for $3\le L\le8$ 
using the Bulirsch-Stoer (BST) algorithm\cite{bst}.
The error estimates are twice the difference
between the ($n-1$,1) and ($n-1$,2) approximants\cite{bst}.
The critical points for $x=1$ (three-state) and $x=0$ (two-state) 
Potts models are known exactly and are included in Table I for comparison.
Note that the imaginary parts of $y_c$(BST) are all consistent with zero.
We have also calculated the thermal exponent, $y_t$, applying
the BST algorithm to the values given by Eq. (11), and these results 
are also presented in Table I. For $x=1$ we find $y_t$
very close to the known value $y_t=6/5$ for the three-state model,
but for $x$ as large as 0.5 we obtain $y_t=1$, the value of the 
thermal exponent for the two-state (Ising) model. 

Figure 2 shows the critical line of the three-state Potts 
ferromagnet for $H_q < 0$.
In Figure 2 
the upper line is the critical temperature of the two-state model, 
$T_c(Q=2)=1/{\rm ln}(1+\sqrt{2})$, 
and the lower line is the critical temperature for the three-state model,
$T_c(Q=3)=1/{\rm ln}(1+\sqrt{3})$.
The critical line for small $-H_q$ is given by\cite{cfp}
\be
T-T_c(Q=3)\sim(-H_q)^{y_t/y_h},
\ee
where $y_t=6/5$ and $y_h=28/15$ for the three-state Potts model.


\section{Fisher zeros of the three-state Potts model for $\lowercase{x}>1$}

In the limit $H_q\to\infty$ ($x\to\infty$) the positive field $H_q$
favors the state $q$ for every site and the $Q$-state Potts model 
is transformed into the one-state model\cite{martin2}.
The zeros are given by
\be
Z_Q\sim y^{-N_b}\sum_{E=0}^{N_b}\Omega_Q(E,M=N_s)y^E=0.
\ee
Because $\Omega_Q(E,M=N_s)=1$ for $E=0$ and 0 otherwise,
Eq. (13) is
\be
y^{-N_b}=0.
\ee
As $x$ increases $|y_0|$ for all the zeros increases without bound.

Figure 3 shows the Fisher zeros in the complex $y$-plane of the 
three-state Potts model for $h=\beta H$ varying from 0 to 4 in steps of 1.
As $h$ increases, all the Fisher zeros move away from the origin.
Note that for $h > 0$ there is accumulation of the Fisher zeros 
as we approach the Fisher edge zero, 
that is, the Fisher zero closest to the positive real axis. 
That kind of accumulation suggests that for $h > 0$ the density of zeros
diverges at the Fisher edge zero, which we call the Fisher edge
{\it singularity}. The critical exponents associated with the edge
singularity are defined in the usual way,
\be
C_e\sim(1-{y\over y_e})^{-\alpha_e},
\ee
\be
m_e\sim(1-{y\over y_e})^{\beta_e},
\ee
and
\be
\chi_e\sim(1-{y\over y_e})^{-\gamma_e},
\ee
where $y_e$ is the location of the Fisher edge singularity, and
$C_e$, $m_e$ and $\chi_e$ are the singular parts of the specific heat,
magnetization, and susceptibility, respectively.

To study the critical behavior at the Fisher edge singularity
we have used the high-field, low-temperature series expansion
for the three-state Potts model due to Enting\cite{enting1,enting2},
which is coded as partial generating functions. 
Table II shows estimates for $y_e$ and $\beta_e$ 
from Dlog Pad\'e approximants\cite{guttmann} for the magnetization
at $x=100$. For this value of $x$ we find $\alpha_e=1.22(2)$,
$\beta_e=-0.197(6)$, and  $\gamma_e=1.20(3)$. 
Note that both $\beta_e$ and $\alpha_e$ are unphysical in that
$\beta_e<0$ implies a divergent magnetization and $\alpha_e>1$
implies a divergent energy density.
The density of zeros near the Fisher edge singularity in the complex
temperature plane is given by\cite{fisher2}
\be
g(y)\sim(1-{y\over y_e})^{1-\alpha_e}.
\ee
Therefore, $\alpha_e > 1$ means that the density of zeros diverges
at the Fisher edge singularity. From $\alpha_e$, $\beta_e$, and
$\gamma_e$ we obtain
\be
\alpha_e+2\beta_e+\gamma_e=2.03(4),
\ee
so that the Rushbrooke scaling law $\alpha+2\beta+\gamma=2$,
which is known to hold at a physical critical point, is also satisfied 
at the Fisher edge singularity. 
From the series expansions for the specific heat,
magnetization, and susceptibility we have obtained the location 
of the Fisher edge singularity
\be
y_e({\rm series})=1.232(2)+1.048(3)i,
\ee
which is in excellent agreement with the value we calculate
by extrapolation from finite-size systems using the BST algorithm,
\be
y_e({\rm BST})=1.233(7)+1.050(3)i.
\ee

We have also studied the critical behavior at the Fisher edge
singularity for several values of $x$. Table III shows the edge
critical exponents and the location of the Fisher edge singularity
for $x=20$, 100 and 200. The edge critical exponents for any $x$
satisfy the relation $\alpha_e+2\beta_e+\gamma_e=2$ within 
our error estimates.
The locations of the Fisher edge singularity obtained from the series
analysis agree very well with those extrapolated from
finite size data by the BST algorithm.
Table III suggests that the values of the edge critical exponents
are independent of $x$.  


\section{Fisher zeros of the $Q > 3$ Potts models 
for nonzero magnetic field}

Using the high-field, low-temperature series expansion of the $Q$-state
Potts model for $4\le Q\le6$\cite{enting2}, we have studied the 
critical behavior at the Fisher edge singularity for $Q > 3$.
Table IV shows the edge critical exponents and the locations of the
Fisher edge singularities for $4\le Q\le6$ and $x=100$.
The edge critical exponents for any $Q$ satisfy the relation
$\alpha_e+2\beta_e+\gamma_e=2$ within our error estimates.
As $Q$ increases $\beta_e$ appears to decrease slightly, 
while $\alpha_e$ and $\gamma_e$ are constant within error.
However, because the uncertainties in $\alpha_e$ and $\gamma_e$ are large,
we do not know whether $\alpha_e$ and $\gamma_e$ are truly
independent of $Q$. 
Even though the Yang-Lee edge singularities have never been studied
for the two-dimensional $Q>2$ Potts models, according to the study of other
models\cite{kortman,fisher1,kurtze1,uzelac} and conformal field
theory\cite{cardy} one expects the critical behavior of the Yang-Lee edge
singularities in two dimensions to be universal.
However, in a study\cite{shrock2} of the Fisher (or complex-temperature)
singularities of the Potts model in the absence of an external
magnetic field Shrock {\it et al.} have observed 
a dependence of the edge critical 
exponents on $Q$. In Table IV the BST estimates and the series
results for the location of the Fisher edge singularities agree with
each other for $Q=4$ and 5. For $Q=6$ we have calculated $\Omega_Q(E,M)$
up to $L=5$, and the BST extrapolation is unreliable
because the maximum size of the lattice is small.
Figure 4 shows the Fisher zeros in the complex
$y$-plane of the six-state Potts model for $x=100$, 
and the location of the edge singularities calculated from the series,
which has been the traditional method\cite{shrock2,matveev}
in the study of the Fisher (or complex-temperature) singularities.


\section{conclusion}

We have studied the Fisher zeros in the complex $y$-plane 
of the $Q$-state Potts model for $x\ne1$ using the microcanonical
transfer matrix and the high-field, low-temperature series expansion.
We have discussed the transformation of the Fisher zeros of 
the $Q$-state Potts model into those of the ($Q-1$)-state Potts
model for $x<1$, and into those of the one-state Potts model
for $x>1$. For $x<1$ we have obtained the critical line
and calculated the critical exponents for several values of $x$. 
From the high-field, low temperature
series expansion we have shown that for $3\le Q\le6$
the specific heat, magnetization,
susceptibility, and the density of zeros diverge algebraically 
at the Fisher edge singularity with characteristic edge exponents 
$\alpha_e$, $\beta_e$, and $\gamma_e$. 




\begin{table}
\caption{The critical temperature $y_c$ and the critical exponent $y_t$ 
of the three-state Potts model for $0\le x\le1$.}
\begin{tabular}{ccclc}
$x$ &$y_c$ (BST) &$y_c$ (exact) &$y_t$ (BST) &$y_t$ (exact) \\
\hline
0     &$0.414(3)+0.0002(4)i$ &0.414213... &1.001(2) &1   \\
0.001 &$0.414(3)+0.0002(4)i$ &            &1.001(2)      \\
0.05  &$0.413(5)+0.0002(3)i$ &            &1.0009(6)     \\
0.5   &$0.400(2)+0.000(2)i$  &            &0.982(21)     \\
1     &$0.366(2)+0.0002(5)i$ &0.366025... &1.195(3) &$6\over5$ \\
\end{tabular}
\end{table}

\begin{table}
\caption{Values of $\beta_e$ and $y_e$ estimated from
Dlog Pad\'e approximants to magnetization for $Q=3$ and $x=100$.}
\begin{tabular}{rcc}
$[N/D]$ &$\beta_e$ &$y_e$ \\
\hline
$[7/8]$   &$-0.2077$  &$1.243256+1.052564i$ \\
$[8/8]$   &$-0.1967$  &$1.236604+1.045093i$ \\
$[9/8]$   &$-0.1952$  &$1.237445+1.043712i$ \\
$[8/9]$   &$-0.1954$  &$1.238006+1.043638i$ \\
$[9/9]$   &$-0.1957$  &$1.236733+1.044346i$ \\
$[10/9]$  &$-0.1939$  &$1.235093+1.044025i$ \\
$[9/10]$  &$-0.1956$  &$1.235556+1.044725i$ \\
$[10/10]$ &$-0.1953$  &$1.235416+1.044633i$ \\
$[11/10]$ &$-0.1930$  &$1.233087+1.045059i$ \\
$[10/11]$ &$-0.1956$  &$1.235523+1.044727i$ \\
$[11/11]$ &$-0.2003$  &$1.233911+1.047358i$ \\
$[12/11]$ &$-0.1975$  &$1.233556+1.046650i$ \\
$[11/12]$ &$-0.1972$  &$1.233675+1.046438i$ \\
$[12/12]$ &$-0.1977$  &$1.233538+1.046729i$ \\
$[13/12]$ &$-0.1975$  &$1.233549+1.046638i$ \\
$[12/13]$ &$-0.1959$  &$1.233478+1.046087i$ \\
$[13/13]$ &$-0.1991$  &$1.233249+1.047366i$ \\
$[14/13]$ &$-0.1966$  &$1.233340+1.046474i$ \\
$[13/14]$ &$-0.1943$  &$1.232810+1.046075i$ \\
$[14/14]$ &$-0.1937$  &$1.231850+1.046918i$ \\
$[15/14]$ &$-0.1970$  &$1.233458+1.046527i$ \\
$[14/15]$ &$-0.1945$  &$1.232889+1.046035i$ \\
$[15/15]$ &$-0.1945$  &$1.231888+1.047107i$ \\
$[16/15]$ &$-0.1966$  &$1.233328+1.046471i$ \\
\end{tabular} 
\end{table}
\begin{table}
\caption{The location and the edge critical exponents of the Fisher
edge singularity for the three-state Potts model.}
\begin{tabular}{cccccc}
$x$ &$y_e$ (BST) &$y_e$ (series) &$\alpha_e$ &$\beta_e$ &$\gamma_e$ \\
\hline
20  &$0.880(2)+0.643(2)i$ &$0.880(4)+0.641(3)i$ &1.2(1)  &$-0.19(1)$  &1.2(1)  \\
100 &$1.233(7)+1.050(3)i$ &$1.232(2)+1.048(3)i$ &1.22(2) &$-0.197(6)$ &1.20(3) \\
200 &$1.436(8)+1.270(1)i$ &$1.436(2)+1.268(2)i$ &1.22(2) &$-0.196(7)$ &1.21(3) \\
\end{tabular}
\end{table}

\begin{table}
\caption{The location and the edge critical exponents of the Fisher
edge singularity for the $4\le Q\le6$ Potts models and $x=100$.}
\begin{tabular}{cccccc}
$Q$ &$y_e$ (BST) &$y_e$ (series) &$\alpha_e$ &$\beta_e$ &$\gamma_e$ \\
\hline
4 &$1.13(6)+0.96(4)i$  &$1.159(5)+0.93(1)i$  &1.18(8) &$-0.180(4)$ &1.12(8) \\
5 &$1.09(4)+0.861(8)i$ &$1.103(4)+0.86(1)i$  &1.2(1)  &$-0.173(4)$ &1.1(1) \\
6 &                    &$1.053(8)+0.811(4)i$ &1.2(1)  &$-0.164(9)$ &1.1(1) \\
\end{tabular}
\end{table}


\begin{figure}
\epsfbox{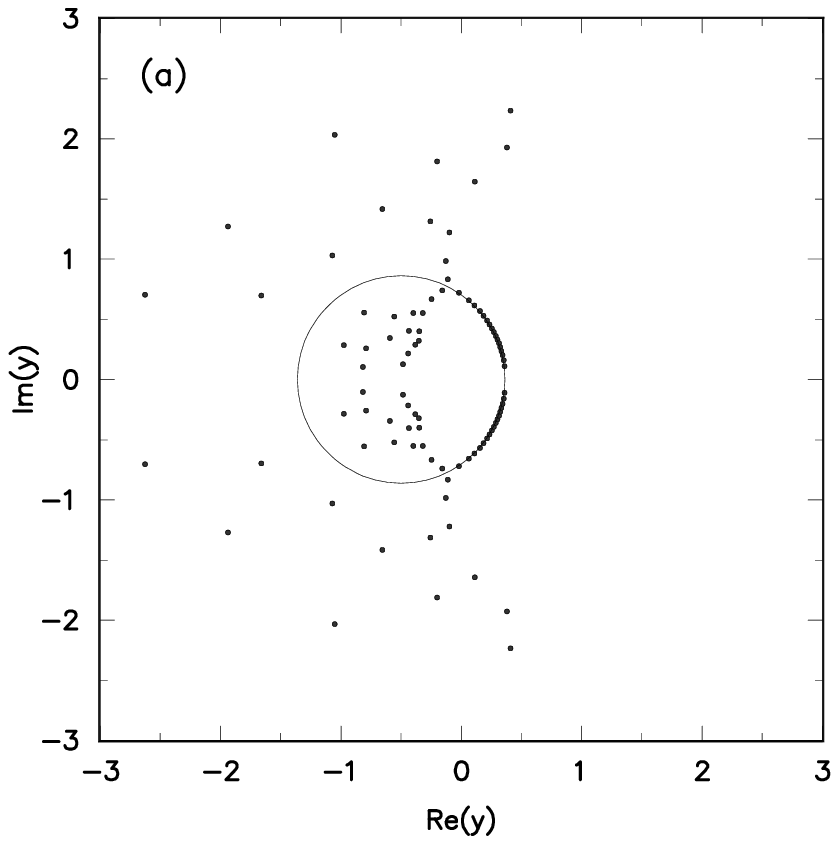}
\epsfbox{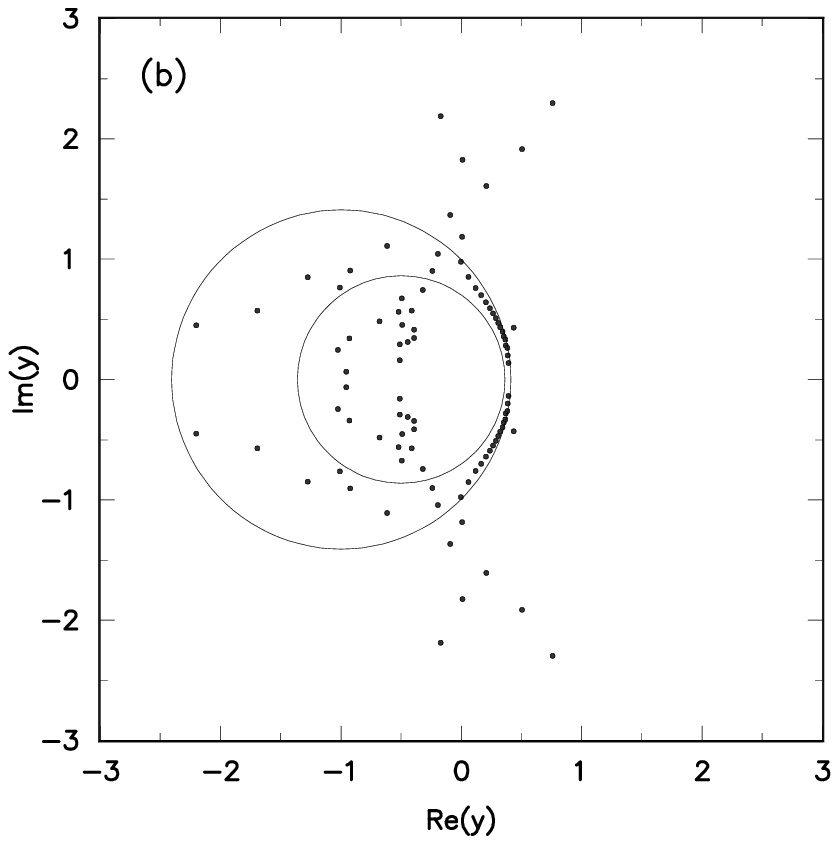}
\epsfbox{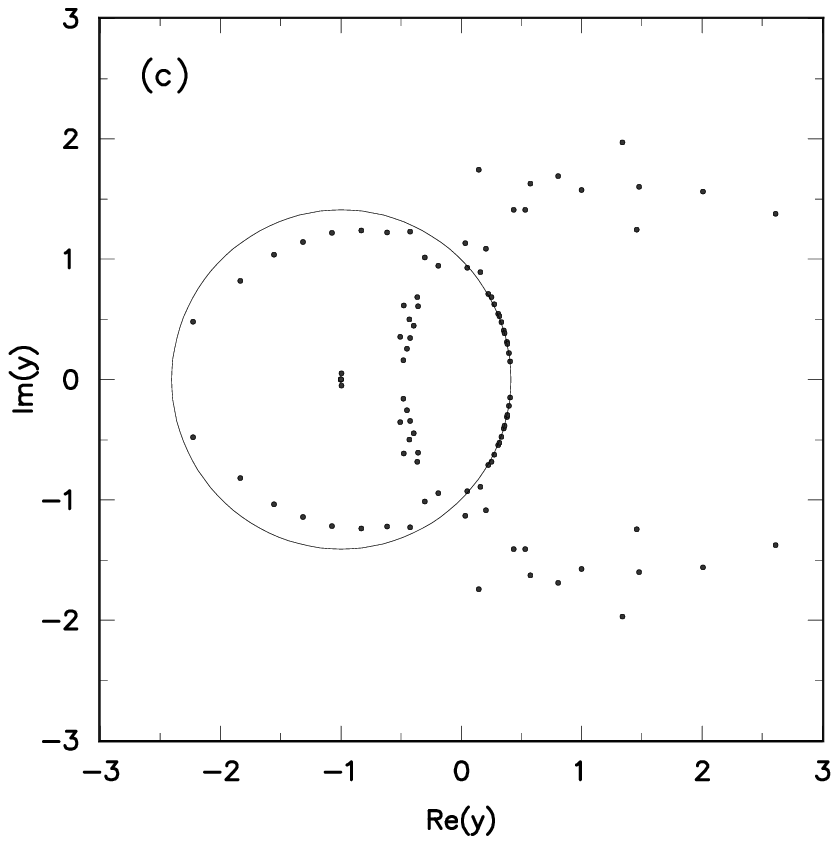}
\epsfbox{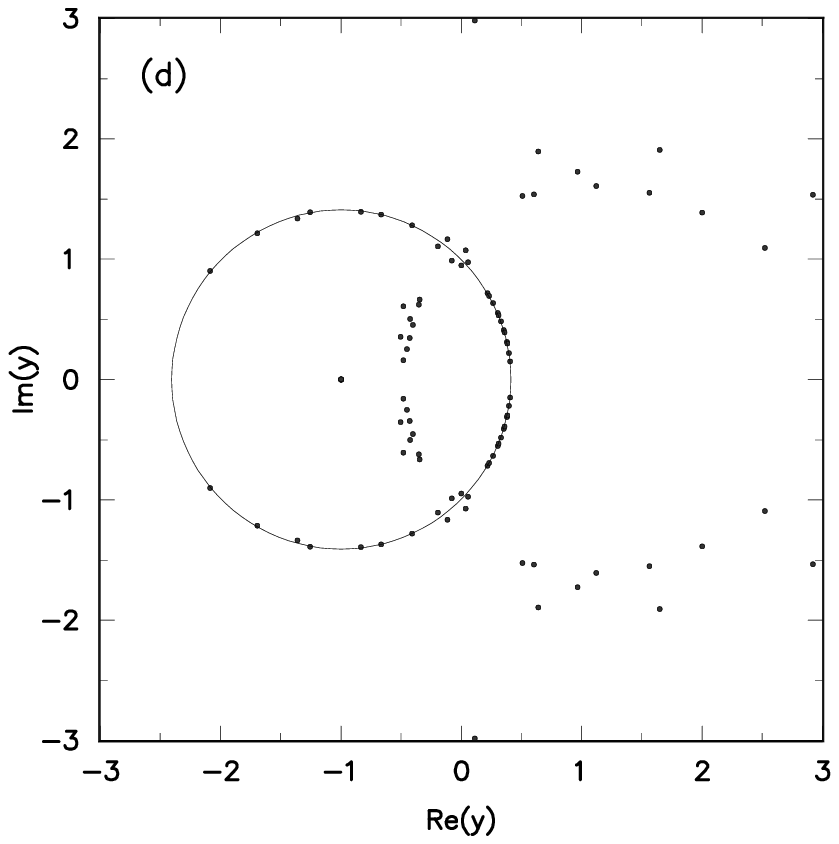}
\caption{Fisher zeros in the complex $y$-plane of a $7\times7$ three-state
Potts model for (a) $x=1$, (b) 0.5, (c) 0.05, and (d) 0.001
with self-dual boundary conditions.}
\end{figure}

\begin{figure}
\epsfbox{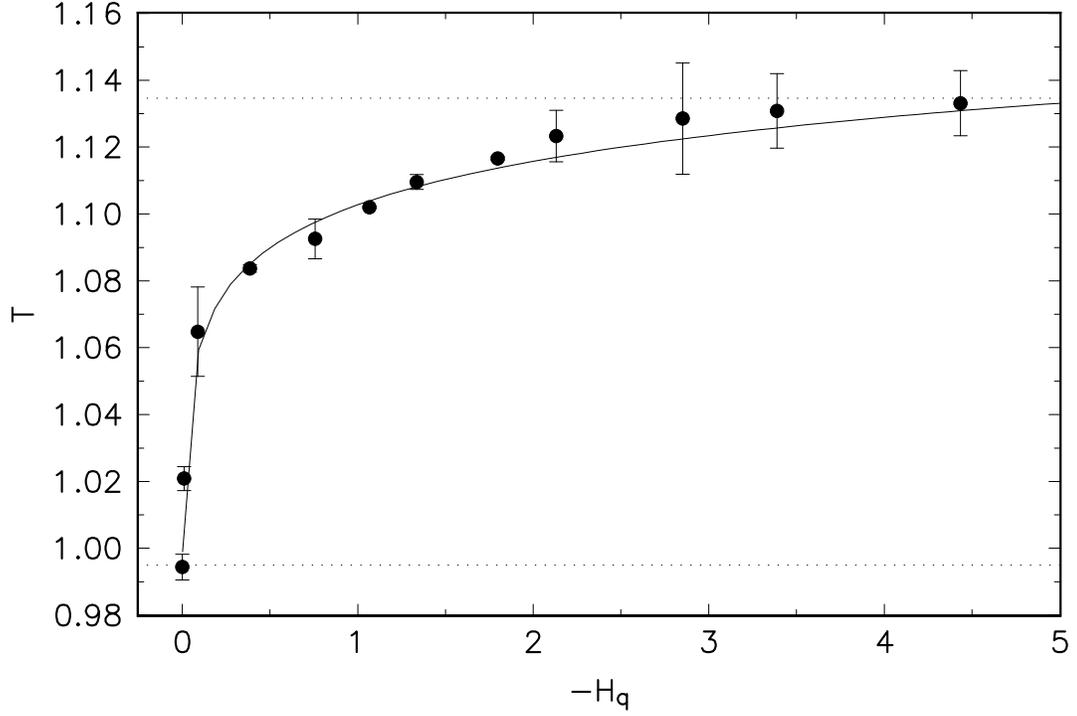}
\caption{Critical temperatures of the three-state Potts ferromagnet
as a function of the magnetic field. $H_q$ is in unit of $J$
and $T$ is in unit of $J/k_B$. The upper dotted line is the Ising
transition temperature in the limit $H_q\to-\infty$, while the lower
dotted line shows the critical temperature of the three-state Potts
model for $H_q=0$.}
\end{figure} 

\begin{figure}
\epsfbox{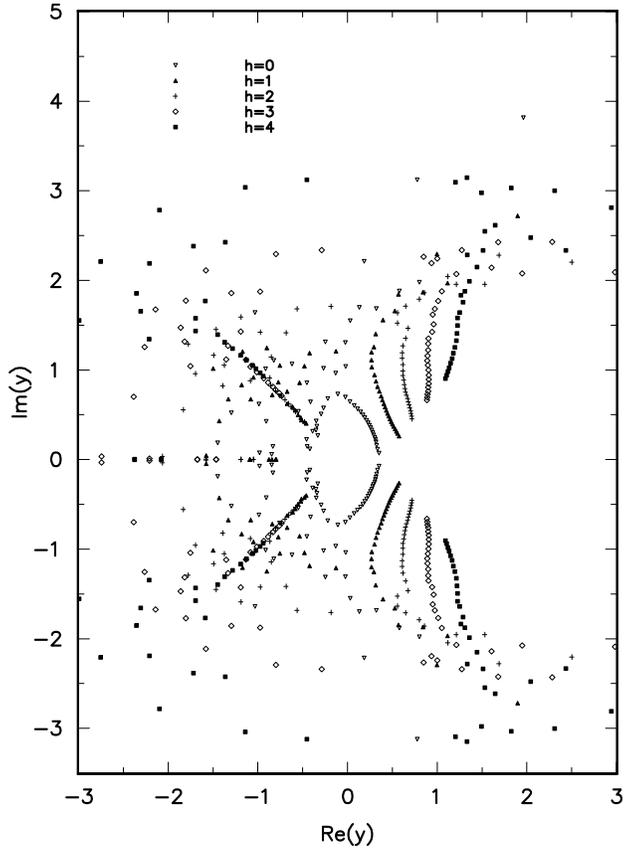}
\caption{Fisher zeros in the complex $y$-plane of an $8\times8$ three-state
Potts model for $h=\beta H$ varying from 0 to 4 in steps of 1 
with cylindrical boundary conditions.}
\end{figure}

\begin{figure}
\epsfbox{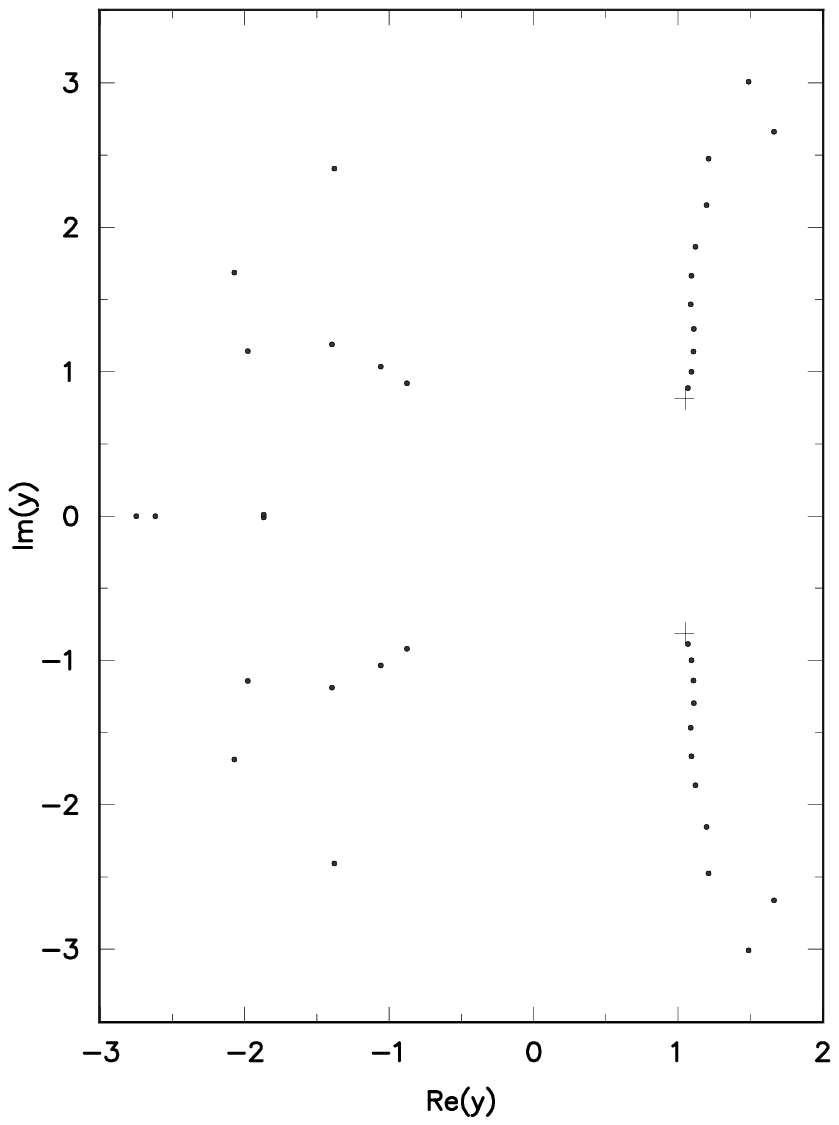}
\caption{Fisher zeros in the complex $y$-plane of a $5\times5$ six-state 
Potts model for $x=100$ with cylindrical boundary conditions.
The two plus symbols show the locations of the Fisher edge singularities
estimated from the series analysis.}
\end{figure} 



\begin{references}
\bibitem{enting2} I. G. Enting, J. Phys. A {\bf7}, 2181 (1974).
In this paper there are two typographical errors. 
In Eq. (17) ($q-2$) should be ($q-3$), and
in the generating function $F_4$ for the six-state Potts model
80(9,4,2,2,0,0,0,1) should be 80(9,4,2,2,0,0,0,0,1).
\end{references}
\end{document}